\documentclass[graybox]{svmult}

\usepackage{mathptmx}       
\usepackage{helvet}         
\usepackage{courier}        
\usepackage{type1cm}        
\usepackage{makeidx}         
\usepackage{graphicx}        
\usepackage{multicol}        
\usepackage[bottom]{footmisc}

\makeindex             

\begin{document}

\title*{ Casimir Effect for the Piecewise Uniform String}
\author{Iver Brevik }
\institute{Iver Brevik \at Department of Energy and Process Engineering, Norwegian University of Science and Technology, N-7491 Trondheim, Norway.
 \email{iver.h.brevik@ntnu.no}}

%
%
\maketitle

\abstract{
The Casimir energy for the transverse oscillations of a piecewise uniform closed string is calculated. In its simplest version the string consists of two parts I and II having in general different tension and mass density, but is always obeying the condition that the velocity of sound is equal to the velocity of light. The model, first introduced by Brevik and Nielsen in 1990, possesses attractive formal properties implying that it becomes easily regularizable by several methods, the most powerful one being the contour integration method.  We also consider the case where the string is divided into $2N$ pieces, of alternating type-I and type-II material. The free energy at finite temperature, as well as the Hagedorn temperature, are found. Finally, we make some remarks on the relationship between this kind of theory and the theory of quantum star graphs, recently considered by Fulling {\it et al.}.
}

\section{Introduction }
\label{sec:1}

Standard theory of closed strings - whatever the string is situated in Minkowski space or in superspace - assumes the string to be homogeneous, i.e. that the tension $T$ is the same everywhere. The composite string model, in which the string is taken to consist of two or more separately different pieces, is a generalization of the usual model. An important condition that we will impose, is that the composite string is relativistic in the sense that the velocity $v_s$ of transverse sound is everywhere assumed to be equal to the velocity of light,
\begin{equation}
v_s=\sqrt{T/\rho}=c=1. \label{1}
\end{equation}
Here $T$, as well as the mass density  $\rho$, refer to the string piece under consideration. At each junction there are two boundary conditions, namely (i) the transverse displacement $\psi=\psi(\sigma, \tau)$ is continuous, and (ii) the transverse force $T\partial \psi/\partial \sigma$ is continuous. Using the equation of motion
\begin{equation}
\left(\frac{\partial^2}{\partial \sigma^2}-\frac{\partial^2}{\partial \tau^2}\right)\psi=0 \label{2}
\end{equation}
one can calculate the eigenvalue spectrum and the Casimir energy of the string.

The simplest string model of this type is when there are only two pieces, of length $L_I$ and $L_{II}$, such that the total length is $L=L_I+L_{II}$; see Fig.~1.
\begin{figure}[htbp]
\makebox[\textwidth][c]{\includegraphics[width=5cm]{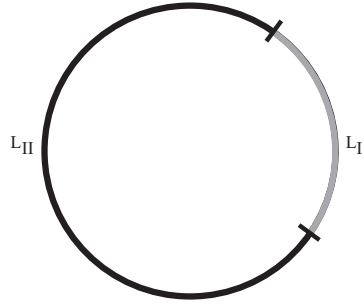}}
\begin{center}
\end{center}
\caption{The two-piece string, with piece lengths $L_I$ and $L_{II}$.}
\end{figure}
This model was introduced in 1990 \cite{brevik90}; cf. also the related paper \cite{li91}. The Casimir energy was calculated for various length ratios of the pieces. It is convenient to introduce a symbol $s$ for the length ratio, and also a symbol $x$ for the tension ratio,
\begin{equation}
s=\frac{L_{II}}{L_I}, \quad x=\frac{T_I}{T_{II}}. \label{3}
\end{equation}
With moreover the function $F(x)$ defined as
\begin{equation}
F(x)=\frac{4x}{(1-x)^2}, \label{4}
\end{equation}
the dispersion relation becomes
\begin{equation}
F(x)\sin^2\left(\frac{\omega L}{2}\right)+\sin \omega L_I\sin \omega L_{II}=0, \label{5}
\end{equation}
and the Casimir energy, describing the deviation from homogeneity,  can be written  formally as
\begin{equation}
E=E_{I+II}-E_{\rm uniform}=\frac{1}{2}\sum \omega_n-E_{\rm uniform}. \label{6}
\end{equation}
Since Eq.~(\ref{5}) is invariant under the substitution $x\rightarrow 1/x$, we can simply assume $x\leq 1$ in the following.

From a physical point of view, there is well-founded hope that
this simple string model can help us to understand the issue of
the energy of the vacuum state in two-dimensional quantum field
theories in general. The system is strikingly easy to regularize,
this  being due to the relativistic property of the model. The
mentioned paper \cite{brevik90} made use of a cutoff
regularization method, whereby a function $f=\exp(-\alpha \omega)$
with $\alpha$ a small positive parameter was introduced. A second
regularization method - the one to be dealt with in this paper -
is the complex contour integration method. To our knowledge this
method was first applied to the composite string model by Brevik
and Elizalde \cite{brevik94}. A separate chapter is devoted to
this model in Elizalde's book on zeta functions \cite{elizalde95}.  The great
advantage of this method is that the {\it multiplicities} of the
zeros of the dispersion function are automatically taken care of.
There exists also a third convenient regularization method
implying the use of the Hurwitz zeta function.

Instead of assuming only two pieces in the composite string, one can imagine that the string is composed of $2N$ pieces, all of the same length, such that the type I materials and the type II materials are alternating. Maintaining the same relativistic property as before, one will find  that also this kind of system is easily regularizable and tractable analytically in general.  There are by now several papers devoted to the study of the composite string in its various facets; cf. Refs.~\cite{brevik95,brevik96,bayin96,brevik97,berntsen97,brevik98,brevik99,brevik03} (the last of these references gives a review).  As for possible applications of the model, we may also mention the paper of Lu and Huang \cite{lu98}, discussing the Casimir energy for a composite  Green-Schwarz superstring.

In the following we review briefly the main properties of the composite string model, at zero, and also at finite, temperature, making use of the contour regularization method as mentioned. The convenience of the recursion formula in the $2N$-case is in our opinion worth attention. The quantum theory of the two-piece string for the simplifying limiting case of very small tension ratio $x$ between the two pieces is highlighted, and the Hagedorn temperature is given for this kind of model. Finally, we comment upon the connection between the theory of the piecewise uniform string and the theory  of quantum star graphs, recently developed by Fulling and others.

\section{ The Two-Piece String}

According to the argument principle one has for any meromorphic function $g(\omega$:
\begin{equation}
 \frac{1}{2\pi i}\oint \omega \frac{d}{d\omega}\ln
g(\omega)d\omega=\sum \omega_0-\sum \omega_\infty, \label{7}
\end{equation}
where $\omega_0$ are the zeros and $\omega_\infty$ are the poles of $g(\omega)$ inside the integration contour. As usual the contour is a semicircle of large radius $R$ in the right half complex $\omega$ plane, closed by a straight line from $\omega=iR$ to $\omega=-iR$.
A convenient choice for the dispersion function $g(\omega)$ is
\begin{equation}
g(\omega)=\frac{F(x)\sin^2[(s+1)\omega L_I/2]+\sin (\omega
L_I)\sin(s\omega L_I)}{F(x)+1}.  \label{8}
\end{equation}
The final result at zero temperature becomes \cite{brevik94}
\begin{equation}
E=\frac{1}{2\pi}\int_0^\infty \ln \left| \frac{F(x)+\frac{\sinh (\xi L_I)\sinh (s\xi L_I)}{\sinh^2[(s+1)\xi L_I/2]}}{F(x)+1} \right| d\xi, \label{9}
\end{equation}
with $\omega=i\xi$. This expression holds for all values of $s$, not necessarily integers. Since it is invariant under the interchange $s \rightarrow 1/s$, we can consider only the interval $s \geq 1$ without any loss of generality. If the tension ratio $x\rightarrow 0$, we find the simple formula
\begin{equation}
E=-\frac{\pi}{24L}\left(s+\frac{1}{s}-2\right). \label{10}
\end{equation}
At finite temperatures, where $\xi_n=2\pi nT$ with $n=0,1,2,3..$ are the Matsubara frequencies, we get the corresponding expression
\begin{equation}
 E(T)=T{\sum_{n=0}^\infty}'\ln
\left|
\frac{F(x)+\frac{\sinh(\xi_nL_I)\sinh(s\xi_nL_I)}{\sinh^2[(s+1)\xi_nL_I/2]}}{F(x)+1}
\right|, \label{11}
\end{equation}
where the prime means that the case $n=0$ is counted with half weight.

We may define two characteristic frequencies in the problem: (i) the thermal frequency $ \omega_T= T= \xi_1/(2\pi)$, and (ii)
 the geometric frequency $\omega_{\rm geom}=2\pi/L_I$. The case of high temperatures corresponds to $ \frac{\omega_T}{\omega_{\rm geom}} \geq 1$, whereby we can approximate
\begin{equation}
 E(T)=\frac{1}{2}T\ln \left| \frac{F(x)+4s/(s+1)^2}{F(x)+1} \right|. \label{12}
 \end{equation}
Thus, if  "our" universe (I) is small and the "mirror" universe (II) is large ($s\rightarrow \infty$), we have
\begin{equation}
 E(T)=-\frac{1}{2}\ln \left|1+F(x)^{-1}\right|. \label{13}
 \end{equation}
In the case of low temperatures, $ \frac{\omega_T}{\omega_{geom}} \ll 1$, a  large number of Matsubara frequencies becomes necessary.

\section{The $2N$-Piece String}

Assume now that the string of length $L$ is divided into $2N$ pieces of equal length, of alternating type I/type II material. The basic formalism for arbitrary integers $N$ was set up in Ref.~\cite{brevik95}, although a full calculation was not worked out until Refs.~\cite{brevik97,berntsen97}. A key point in \cite{brevik97} was the derivation of a new recursion formula for the matrix of the dispersion function, applicable for general integers $N$.

In addition to the tension ratio $x=T_I/T_{II}$ we define two new symbols,
\begin{equation}
 p_N=\frac{\omega L}{N}, \quad \alpha=\frac{1-x}{1+x}. \label{14}
\end{equation}
The eigenfrequencies are determined from
\begin{equation}
 \rm{Det}\left[ {\bf M}_{2N}(x,p_N)-{\bf 1} \right]=0, \label{15}
 \end{equation}
 where  the system determinant satisfies the following recursion relation
 \begin{equation}
 {\bf M}_{2N}(x,p_N)=\left[ \frac{(1+x)^2}{4x}\right]^N { \Lambda}^N(\alpha, p_N), \label{16}
\end{equation}
with
\begin{equation}
  \Lambda (\alpha, p)=\left( \begin{array}{ll}
a & b \\
b^* & a^*
\end{array}
\right), \label{17}
\end{equation}
\begin{equation}
 a=e^{-ip}-\alpha^2, \quad b=\alpha (e^{-ip}-1). \label{18}
 \end{equation}
This property greatly facilitates the handling of the formalism.
The way to proceed now is to calculate the eigenvalues of the
matrix $\Lambda$, and express the elements of ${\bf M}_{2N}$ as
powers of these. One finds
\begin{equation}
\lambda_\pm(iq)=\cosh q-\alpha^2 \pm [(\cosh
q-\alpha^2)^2-(1-\alpha^2)^2]^{1/2}, \label{19}
\end{equation}
where $\lambda_\pm$ are the eigenvalues of $\Lambda$ for imaginary
arguments $iq$ of the dispersion equation.

The contour integration method gives for the  Casimir energy
($T=0$):
\begin{equation}
 E_N(x)=\frac{N}{2\pi L}\int_0^\infty \ln \left|
\frac{2(1-\alpha^2)^N-[\lambda_+^N(iq)+\lambda_-^N(iq)]}{4\sinh^2(Nq/2)}
\right| dq. \label{20}
\end{equation}
It is seen that $E_N(x)<0,~|E_N(x)|$ increasing with increasing
$N$. Division into a larger number of pieces thus diminishes the
Casimir energy.

If $x\rightarrow 0$,
\begin{equation}
 E_N(0)=-\frac{\pi}{6L}(N^2-1). \label{21}
 \end{equation}

  We could alternatively use zeta function regularization here.
  That would necessitate, however, solution of the eigenvalue
  spectrum. Degeneracies would have to be put in by hand. The
  latter method is therefore most convenient for low integers $N$.

A rather unexpected property of the system is that of {\it scaling
invariance}. This is seen by examining the behavior of the
function $f_N(x)$ defined by
\begin{equation}
 f_N(x)=\frac{E_N(x)}{E_N(0)}, \quad 0<f_N(x)<1. \label{22}
 \end{equation}
Numerically it turns out that the curve for $f_N(x)$ is
practically the same, irrespective of the value of $N$, as long as
$N\geq 2$. The simple analytic form
 \begin{equation}
 f_N(x) \rightarrow f(x)=(1-\sqrt{x})^{5/2}
 \label{23}
 \end{equation}
fits the numerical values accurately, in particular in the region
$0<x<0.45$. The reason for this behavior is not known.

At finite temperature the expression for the Casimir energy
becomes
\begin{equation}
E_N^T(x)=T{\sum_{n=0}^\infty}' \ln \left|
\frac{2(1-\alpha^2)^N-[\lambda_+^N(i\xi_n L/N)+\lambda_-^N(i\xi_n
L/N)]}{4\sinh^2(\xi_nL/2)}\right|, \label{24}
\end{equation}
where $\lambda_\pm (i\xi_nL/N)$ are given by Eq.~(\ref{19}) with
$q\rightarrow q_n=\xi_nL/N$. It is here useful to note that
\begin{equation}
\lambda_+(iq_n)+\lambda_-(iq_n)=2(\cosh q_n-\alpha^2).
\label{25}
\end{equation}
There are several special cases of interest. First, if the string
is uniform ($x=1$), we get $E_N^T(1)=0$. This is as expected, as
the Casimir energy is a measure of the string's inhomogeneity. If
$N=1$, $x$ arbitrary, we also get a vanishing result,
$E_1^T(x)=0$. In particular, if $x\rightarrow 0$ we get the simple
formula
\begin{equation}
E_N(0)=2T{\sum_{n=0}^\infty }' \ln
\left|\frac{2^N\sinh^N(\xi_nL/2N)}{2\sinh(\xi_nL/2)}\right|.
\label{26}
\end{equation}

\section{  Oscillations of the Two-Piece String in  $D$-Dimensional Spacetime. Quantization}

We will now aim at sketching the essentials of the quantum theory of the composite string, in the case when  $N=1$ (the two-piece string). To allow for a correspondence to the superstring, we allow the number of flat spacetime dimensions $D$ to be an arbitrary integer. In accordance with usual practice, we put now  $L=L_I+L_{II}=\pi$. The theory will be based on two simplifying assumptions:

(i)  The string tension ratio $x\rightarrow 0$. The dispersion relation (\ref{5}) leads in this case to two different branches of solutions, namely the first branch obeying
\begin{equation}
\omega_n(s)=(1+s)n, \label{27}
\end{equation}
and the second branch obeying
\begin{equation}
\omega_n(s^{-1})=(1+s^{-1})n, \label{28}
\end{equation}
with $n=\pm 1, \pm 2, \pm 3,....$

(ii) The second assumption is that the length ratio $s$ is an integer, $s=1,2,3,...$.

\bigskip

Let now $ X^\mu(\sigma, \tau)$ with $\mu=0,1,2,..(D-1)$ be the coordinates on the world sheet. For each branch
\begin{equation}
 X^\mu=x^\mu+\frac{p^\mu\tau}{\pi \bar{T}(s)}+\theta(L_I-\sigma)X_I^\mu+\theta(\sigma-L_I)X_{II}^\mu, \label{29}
 \end{equation}
 where $\theta(x)$ is the step function, $x^\mu$ the center of mass position, and $p^\mu$ the total  momentum of the string. The mean tension in the actual limit is $ \bar{T}(s)={T_{II}s/(1+s)}$ (we assume $T_{II}$ finite). The string's translational energy is
$ p^0=\pi \bar{T}(s)$. In the following we consider the first branch only.

In region I we make the expansion
\begin{equation}
 X_I^\mu=\frac{i}{2\sqrt{\pi T_I}}\sum_{n\neq
0}\frac{1}{n}\left[\alpha_n^\mu(s)e^{i(1+s)n(\sigma-\tau)}+\tilde{\alpha}_n^\mu(s)e^{-i(1+s)n(\sigma+\tau)}\right],
 \label{30}
 \end{equation}
 where $ \alpha_{-n}^\mu=(\alpha_n^\mu)^*, \,
\tilde{\alpha}_{-n}^\mu=(\tilde{\alpha}_n^\mu)^*$. The action can be expressed as
\begin{equation}
 S=-\frac{1}{2}\int d\tau d\sigma
T(\sigma)\eta^{\alpha\beta}\partial_\alpha X^\mu\partial_\beta
X_\mu, \label{31}
\end{equation}
 where $ T(\sigma)=T_I+(T_{II}-T_I)\theta(\sigma-L_I)$. As the conjugate momentum is $P^\mu(\sigma)=T(\sigma)\dot{X}^\mu,$ we obtain the Hamiltonian
\begin{equation}
H=\int_0^\pi [P_\mu(\sigma)\dot{X}^\mu-L]d\sigma=\frac{1}{2}\int_0^\pi T(\sigma)(\dot{X}^2+{X'}^2)d\sigma.
\label{32}
\end{equation}
The fundamental condition is that  $H=0$ when applied to physical states.

The corresponding expansion of the first branch in
region II is
\begin{equation}
 X_{II}^\mu=\frac{i}{2\sqrt{\pi T_I}}\sum_{n\neq 0}
\frac{1}{n}\gamma_n^\mu(s)e^{-i(1+s)n\tau}\cos [(1+s)n\sigma], \label{33}
\end{equation}
with
\begin{equation}
 \gamma_n^\mu(s)=\alpha_n^\mu(s)+\tilde{\alpha}_n^\mu(s), \quad
n\neq 0. \label{34}
\end{equation}
The condition $x\rightarrow 0 $ means that there are only  standing waves
in region II.

We may now introduce light-cone coordinates $ \sigma^-=\tau-\sigma, \, \sigma^+=\tau+\sigma$. Some calculation shows that the total Hamiltonian can be written as a sum of two parts,
\begin{equation}
 H=H_I+H_{II}, \label{35}
 \end{equation}
where
\begin{equation}
 H_I=\frac{1+s}{4}\sum_{-\infty}^\infty [\alpha_{-n}(s)\cdot
\alpha_n(s)+\tilde{\alpha}_{-n}(s)\cdot \tilde{\alpha}_n(s)], \label{36}
\end{equation}
\begin{equation}
 H_{II}=\frac{s(1+s)}{8x}\sum_{-\infty}^\infty
\gamma_{-n}(s)\cdot \gamma_n(s). \label{37}
\end{equation}
The mass $M$ of the string determined by $M^2=-p^\mu p_\mu$,
\begin{equation}
 M^2=\pi T_{II}s\sum_{n=1}^\infty \left[ \alpha_{-n}(s)\cdot
\alpha_n(s)+\tilde{\alpha}_{-n}(s)\cdot
\tilde{\alpha}_n(s)+\frac{s}{2x}\gamma_{-n}(s)\cdot
\gamma_n(s)\right]. \label{38}
\end{equation}
Recall that this is the contribution from first branch only. The expression is valid
for even/odd values of $s$.

Consider now the quantization of the first branch modes. We impose
the conditions
\begin{equation}
 T_I[ \dot{X}^\mu(\sigma,
\tau),X^\nu(\sigma',\tau)]=-i\delta(\sigma-\sigma')\eta^{\mu\nu}
\label{39}
\end{equation}
in region I, and
\begin{equation}
 T_{II}[ \dot{X}^\mu(\sigma,
\tau),X^\nu(\sigma',\tau)]=-i\delta(\sigma-\sigma')\eta^{\mu\nu}
\label{40}
\end{equation}
in region II (the other commutators vanish). Then introducing
creation and annihilation operators via
\begin{equation}
\alpha_n^\mu(s)=\sqrt{n}a_n^\mu(s), \quad
\alpha_{-n}^\mu(s)=\sqrt{n}{a_n^\mu}^ \dagger (s), \label{41}
\end{equation}
\begin{equation}
 \gamma_n^\mu(s)=\sqrt{4nx}c_n^\mu(s), \quad
\gamma_{-n}(s)=\sqrt{4nx}{c_n^\mu}^\dagger (s), \label{42}
\end{equation}
one arrives at the conventional commutation relations
\begin{equation}
 [a_n^\mu(s), {a_m^\nu}^\dagger (s)]=\delta_{nm}\eta^{\mu\nu},
 \quad
 [c_n^\mu(s),{c_m^\nu}^\dagger (s)]=\delta_{nm}\eta^{\mu\nu},
 \label{43}
 \end{equation}
for $n,m \geq 1$.

Now introduce $t(s)$ as
\begin{equation}
t(s)=\pi \bar{T}(s), \label{44}
\end{equation}
and put $D=26$, the usual dimension for the bosonic string. The
condition  $H=H_I+H_{II}=0 $ leads to
\[
 M^2=t(s)\sum_{i=1}^{24}\sum_{n=1}^\infty
\omega_n(s)[{a_{ni}}^\dagger (s)a_{ni}(s)+{\tilde{a}_{ni}}^\dagger
(s)\tilde{a}_{ni}(s)-2] \]
\begin{equation}
+2st(s)\sum_{i=1}^{24}\sum_{n=1}^\infty \omega_n(s)[
{c_{ni}}^\dagger (s)c_{ni}(s)-1], \label{45}
\end{equation}
and the free energy becomes
\[ F=-\frac{1}{24}(s+\frac{1}{s}-2)-2^{-40}\pi^{-26}t(s)^{-13}\int_0^\infty \frac{d\tau_2}{\tau_2^{14}}\int_{-1/2}^{1/2}d\tau_1 \]
\begin{equation}
 \times \left[ \theta_3\left(0\Big|
\frac{i\beta^2t(s)}{8\pi^2\tau_2}\right)-1\right]\Big|
\eta[(1+s)\tau]\Big|^{-48}\eta[s(1+s)(\tau-\bar{\tau})]^{-24}
\label{46}
\end{equation}
Here $ \tau=\tau_1+i\tau_2$ is the
 $\rm Teichm\ddot{u}ller$ parameter,
 \begin{equation}
 \eta(\tau)=e^{\pi i\tau/12}\prod_{n=1}^\infty \left( 1-e^{2\pi
in\tau}\right)  \label{47}
\end{equation}
is the Dedekind $\eta$-function, and
\begin{equation}
 \theta_3(v|x)=\sum_{n=-\infty}^\infty e^{ixn^2+2\pi i
 vn}\label{48}
 \end{equation}
 is the Jacobi $\theta_3$- function. From this the thermodynamic
 quantities such as internal energy $U$ and entropy $S$ can be
 calculated,
 \begin{equation}
 U=\frac{\partial(\beta F)}{\partial \beta}, \quad
S=\beta^2\frac{\partial F}{\partial \beta}. \label{49}
\end{equation}
Finally, it is of interest to write down the Hagedorn temperature
$ T_c=1/\beta_c$, as the free energy $  F \rightarrow \infty $ for
$T>T_c$. We get
\begin{equation}
 \beta_c=\frac{4}{s}\sqrt{\frac{\pi (1+s)}{T_{II}}} \quad (T_{II}~~\rm
 assumed ~~finite). \label{50}
 \end{equation}
In the point mass limit the formalism simplifies somewhat. For
dimensional reasons we must in that case have
\begin{equation}
  F \propto 1/L_I =(1+s)/\pi \approx s/\pi . \label{51}
  \end{equation}
  Readers interested in a more detailed exposition of this theory
  may consult Refs.~\cite{brevik98,brevik99,brevik99a}.

  \section{Final Remarks}
  The piecewise uniform string model is a natural generalization
  of the conventional uniform string. The adaptability of the
  formalism to various regularization schemes, in particular the
  contour integration method, should be emphasized. Of course, an
  important factor here is the assumption about relativistic
  invariance, as illustrated already by Eq.~(\ref{1}). If this
  assumption were removed, the formalism would be difficult to
  handle.

  Another point worth noticing is the close connection between
  the relativistic invariance property and the theory of an
  electromagnetic field propagating in a so-called isorefractive
  medium meaning that  the refractive index is equal to one, or at least  a constant everywhere in  the material system. Recent
  works in this direction are, for instance, Refs.~\cite{brevik09,ellingsen09}. Again, if the isorefractive (or relativistic) condition were removed, the regularization procedure would be rather difficult to deal with, as the contact term to be subtracted off would then depend on which of the media one chooses for this purpose.

  As a proposal for future work, we mention that there may  be a connection between the phases of the piecewise uniform (super) string and the Bekenstein-Hawking entropy associated with this string. The entropy, as known, can be derived by counting black hole microstates, and it is natural to expect that the deviation from spatial homogeneity infaced in the composite string model could influence that sort of calculations.

  Finally, we mention the interesting analogy that seems to exist between the composite string model and the so-called quantum star graph model. Fulling {\it et al.} \cite{fulling07} recently studied vacuum energy and Casimir forces in one-dimensional quantum graphs (pistons), and found that the piston force could be attractive or repulsive depending on the number of edges. It may be that the mathematical similarities between these two kinds of theories reflect a deeper physical similarity also. This remains to be explored.

\bigskip

  {\bf  Acknowledgment}: I thank Stephen A. Fulling for information about Ref.~\cite{fulling07}.

\end{document}